\documentclass[12pt]{article}
\usepackage{epsfig}
\usepackage{cite}
\input amssym.def
\input amssym

\def\beq{\begin{equation}}
\def\eeq{\end{equation}}
\def\bea{\begin{eqnarray}}
\def\eea{\end{eqnarray}}

\begin{document}

\rightline{CERN-PH-TH/2004-157}
\rightline{November 2004}

\vspace{0.5cm}

\begin{center}
{\Large\bf Numerical analysis}\\
{\Large\bf of the Balitsky-Kovchegov equation}\\ 
{\Large\bf with running coupling:}\\
{\Large\bf dependence of the saturation scale}\\
{\Large\bf on nuclear size and rapidity}\\
\vspace{1cm}
 
J.~L.~Albacete$^{1,2}$, N.~Armesto$^{2}$,
J.~G.~Milhano$^{2,3}$,
C.~A.~Salgado$^{2}$ and U.~A.~Wiedemann$^2$
 
\vspace{0.2cm}
 
{\it $^1$ Departamento de F\'{\i}sica, M\'odulo C2, Planta baja,
Campus de Rabanales, Universidad de C\'ordoba, 14071 C\'ordoba, Spain
} \\
\vskip 0.2cm
{\it $^2$ Department of Physics, CERN, Theory Division,
CH-1211 Gen\`eve 23, Switzerland
}\\
\vskip 0.2cm
{\it $^3$ CENTRA, Instituto Superior T\'ecnico (IST),\\
Av. Rovisco Pais, P-1049-001 Lisboa, Portugal
}
\end{center}

{\small We study the effects of including a running coupling constant in
high-density QCD evolution. For fixed coupling constant, QCD evolution
preserves the initial dependence of the saturation momentum $Q_s$ on 
the nuclear size $A$ and results in an exponential dependence
on rapidity $Y$, $Q^2_s(Y) = Q^2_s(Y_0) \exp{\left[ \bar\alpha_s\, d\, (Y-Y_0)
\right]}$.  For the running coupling case, we re-derive analytical estimates
for the $A$- and $Y$-dependences of the saturation scale and test them
numerically. The $A$-dependence of $Q_s$ vanishes $\propto 1/ \sqrt{Y}$ for
large $A$ and $Y$.  The $Y$-dependence is reduced to $Q_s^2(Y) \propto
\exp{(\Delta^\prime\sqrt{Y+X})}$ where we find numerically
$\Delta^\prime\simeq 3.2$.
We study the behaviour of the gluon distribution at large transverse
momentum, characterizing it by an anomalous dimension $1-\gamma$ which we
define in a fixed region of small dipole sizes.
In contrast to previous analytical work, we find a marked difference between
the fixed coupling
($\gamma \simeq 0.65$) and running coupling ($\gamma \sim 0.85$) results.
Our numerical findings
show that both a scaling function depending only on the variable $r\,Q_s$
and the perturbative double-leading-logarithmic
expression, provide equally good descriptions of the
numerical solutions for very small $r$-values below the so-called scaling
window.
}

\newpage

\section{Introduction} \label{intro}

High-density QCD \cite{cargese} -- the regime of large gluon densities
-- provides an experimentally accessible testing ground for 
our understanding of QCD beyond standard perturbation theory.
The Balitsky-Fadin-Kuraev-Lipatov (BFKL)
equation \cite{Kuraev:1977fs,Balitsky:1978ic}
is the perturbative framework
in which the evolution of parton densities with decreasing Bjorken-$x$
(increasing energy) is usually discussed. In the BFKL equation it is
implicitly assumed that the system remains dilute throughout evolution
and hence correlations between partons can be neglected. The fast
growth of the gluon density predicted by the BFKL equation and
experimentally observed at HERA, eventually leads to a situation in
which individual partons necessarily overlap and, therefore, finite
density effects need to be included in the evolution. These effects
enter the evolution non-linearly, taming the growth of the gluon
density.

The need for and role played by saturation effects was first discussed
in \cite{Gribov:1984tu,Mueller:1985wy}.
It was later argued \cite{McLerran:1993ni,McLerran:1993ka,McLerran:1994vd}
that in the high-density domain a hadronic
object (hadron or nucleus) can be described in terms of an ensemble of
classical gluon fields, and that the number of gluons with momenta smaller
than the so-called saturation scale, is as high as it may be (i.e. saturated).
The quantum evolution of the hadronic ensemble
can be written in terms of a non-linear functional equation
\cite{Jalilian-Marian:1996xn,Jalilian-Marian:1997gr,Jalilian-Marian:1998cb,
Kovner:1999bj,Iancu:2000hn,Iancu:2001ad,Ferreiro:2001qy} where the density 
effects are treated non-perturbatively 
(see also \cite{Buchmuller:1995mr,Ayala:1996em}).

An alternative approach, followed by Balitsky \cite{Balitsky:1995ub},
relies on the operator product expansion for high-energy QCD to derive 
a hierarchy of coupled evolution equations (see \cite{Weigert:2000gi} 
for a more compact derivation). In the limit of a large number of colours, 
the hierarchy reduces to one closed equation. This equation was derived
independently by Kovchegov \cite{Kovchegov:1999yj}
in the dipole model of high-energy
scattering \cite{Nikolaev:1990ja,Mueller:1993rr,Mueller:1994jq}.

The relation between these two approaches has been extensively
discussed \cite{Iancu:2000hn,Iancu:2001ad,Ferreiro:2001qy,
Kovner:2000pt,Mueller:2001uk,Blaizot:2002xy,Iancu:2003uh}.
Apart from possible differences between the evolution equations in the 
kinematical region where the projectile becomes dense \cite{Kovner:2000pt}, 
the different approaches yield the same result, usually known
as the Balitsky-Kovchegov (BK) equation. This equation has served as the 
starting point for a large number of analytical and numerical studies.
It has also been derived in the $S$-matrix approach of \cite{Kovner:2001vi},
and as the large-$N_c$ limit of the sum of fan diagrams of BFKL 
ladders \cite{Braun:2000wr,Bartels:2004ef}. It corresponds, as BFKL,
to a re-summation of the leading terms in $\alpha_s\ln{(s/s_0)}$ (leading-log
approximation).

Although the full analytical solution of the BK equation is not known, 
several of its general properties, such as the existence and form of
limiting solutions, have been 
identified in both analytical \cite{Levin:1999mw,Iancu:2002tr,Mueller:2002zm,
Mueller:2003bz,Munier:2003vc,Munier:2003sj,Munier:2004xu} and numerical
\cite{Braun:2000wr,Kimber:2001nm,Armesto:2001fa,Levin:2001et,Lublinsky:2001bc,
Golec-Biernat:2001if,Albacete:2003iq} studies. Most of them refer to the 
fixed coupling case without impact parameter dependence, but analyses of 
the effect of a running coupling \cite{Lublinsky:2001yi,
Triantafyllopoulos:2002nz,
Golec-Biernat:2001if,Braun:2003un,Rummukainen:2003ns,Kutak:2004ym}
and of the dependence on impact parameter
\cite{Kovner:2001bh,Golec-Biernat:2003ym,Gotsman:2004ra} have also been 
carried out. Besides, there have been attempts to go beyond
the large-$N_c$ limit, either by analytical arguments
\cite{Iancu:2003zr,Mueller:2004se,Iancu:2004es}
or by numerically solving the full
hierarchy of evolution equations \cite{Rummukainen:2003ns}. In this latter 
work, non-leading $N_c$ corrections are found to
give a contribution smaller than $10\div 15$\%, in qualitative agreement
with what could be naively expected from
a numerical correction of ${\cal O}(1/N_c^2)$.
From a phenomenological point of view, studies of the BK equation
are motivated by the geometrical scaling phenomenon
observed in lepton-proton \cite{Stasto:2000er}
and lepton-nucleus data \cite{Freund:2002ux,Armesto:2004ud} which has been
related to the scaling properties of the solution of the BK equation (see e.g.
\cite{Gotsman:2002yy,Eskola:2002yc} for recent numerical analyses of HERA data
based on non-linear evolution).
Further interest comes from the study of nuclear collisions 
\cite{Armesto:2003yf}, where saturation physics is argued 
\cite{Gyulassy:2004zy} to underlie a large body of data including
multiplicity distributions \cite{Armesto:2004ud,
Kharzeev:2000ph,Kharzeev:2001gp,Kharzeev:2004if,Braun:2004qh,Braun:2004kp}
and the rapidity 
dependence of the Cronin effect
\cite{Kharzeev:2002pc,Baier:2003hr,Kharzeev:2003wz,Albacete:2003iq,
Kharzeev:2004yx}. 

Next-to-leading-log contributions \cite{Fadin:1998py,Ciafaloni:1998gs} are known
to have a strong impact on the BFKL equation
\cite{Ross:1998xw,Kovchegov:1998ae,Armesto:1998gt,Andersen:2003an,
Andersen:2003wy}.  Both the
choice of scale in the coupling constant \cite{Brodsky:1998kn} and the
implementation of kinematical cuts for gluon emission
\cite{Schmidt:1999mz,Forshaw:1999xm,Golec-Biernat:2001if}, together with
physically motivated modifications of the kernel
\cite{Ciafaloni:2003kd,Altarelli:2003hk,Khoze:2004hx},
have been proposed to mend some observed pathologies of next-to-leading-log
BFKL.  
It is usually expected that the unitarity corrections included in the BK 
equation become of importance for parametrically smaller rapidities 
\cite{Kovchegov:1998ae,Armesto:1998gt} than
those for which running coupling effects must be included
\cite{Mueller:1996hm}. This can only be definitively established
once next-to-leading-log contributions are fully computed for BK (see 
\cite{Babansky:2002my} for a first step in this direction). However, 
the inclusion of running coupling effects in BK may offer a hint of 
some of the effects induced at next-to-leading-log, as has been 
previously the case for BFKL. It may also
help to reconcile the results of the equation with phenomenology
\cite{Triantafyllopoulos:2002nz,Armesto:2004ud}.

In this paper we investigate numerically the influence of the running coupling
on the solution of the BK equation without impact parameter dependence,
leaving this last point for a future publication.
We go beyond previous numerical studies
\cite{Golec-Biernat:2001if,Braun:2003un,Rummukainen:2003ns}
by making a detailed comparison between analytical estimates and our
numerical solution of the BK equation, and analyzing
the $Y$- and $A$-dependence of 
the saturation scale. Our key results are the confirmation
of the $Y$- and $A$-dependence of the saturation scale proposed
analytically \cite{Iancu:2002tr,Mueller:2002zm,Mueller:2003bz},
and the novel finding that the anomalous dimension (extracted
for dipole sizes smaller than the 
inverse saturation scale) is different in the fixed and running coupling cases.
To compare to
analytical results which have been
derived for asymptotically large energies, we shall
evolve numerically to very large rapidities (up to $Y\sim 80$), significantly 
beyond the experimentally accessible range. 

The plan of the paper is as follows. We first introduce the BK equation 
in Section \ref{sec:bk} and the different implementations of the running 
of the coupling constant in Section \ref{running}. In Section \ref{numerical}
we explain the numerical method used to solve the BK equation. 
In Section \ref{results} we present our numerical results, and we compare 
with previous numerical works and with analytical estimates. Finally, we
summarize and discuss our main conclusions.

\section{The Balitsky-Kovchegov equation} \label{sec:bk}

The BK equation gives the evolution with rapidity 
$Y=\ln{(s/s_0)}=\ln{(x_0/x)}$ of the
scattering
probability $N(\vec{x},\vec{y},Y)$
of a $q\bar q$
dipole with a hadronic target, where $\vec{x}$ ($\vec{y}$) is
the position of the
$q$ ($\bar q$) in transverse space with respect to the center of the target.
We define
\beq
\vec{r}=\vec{x}-\vec{y},\ \ \vec{b}={\vec{x}+\vec{y}\over 2}\,,\ \ 
\vec{r}_1=\vec{x}-\vec{z},\ \ 
\vec{r}_2=\vec{y}-\vec{z}.
\label{coord}
\eeq
If one neglects the impact parameter dependence (which is justified
for $r\ll b$, i.e. an homogeneous target with radius much larger than any
dipole size to be considered),
the BK equation reads ($r\equiv |\vec{r}|$)
\beq
\frac{\partial N(r,Y)}{\partial Y}=\int {d^2z\over 2\pi}\, K(\vec{r},
\vec{r}_1,\vec{r}_2)\left[N(r_1,Y)+N(r_2,Y)-N(r,Y)-N(r_1,Y)N(r_2,Y)\right],
\label{eq:bk}
\eeq
where the BFKL kernel is
\beq
K(\vec{r},
\vec{r}_1,\vec{r}_2)=\bar \alpha_s\,{r^2 \over r_1^2r_2^2}\ \ ,
\ \ \bar \alpha_s={\alpha_sN_c\over \pi}\,.
\label{kernelfc}
\eeq
The coupling constant is fixed and the kernel is conformally invariant. 
This implies that no impact parameter can be generated if not 
present in the initial condition. Also, there is no divergence for
$r_1,r_2\to 0$ provided $N(r,Y)\propto r^\beta$ for $r\to 0$ with $\beta>0$.
This comes from the cancellation between real and virtual
corrections inherited from the BFKL equation. The azimuthally symmetric 
form of
the BFKL equation, which gives the dominant contribution at high 
energies, corresponds to Equation
(\ref{eq:bk}) without the non-linear term.

\begin{figure}[t]
\begin{center}
\epsfig{file=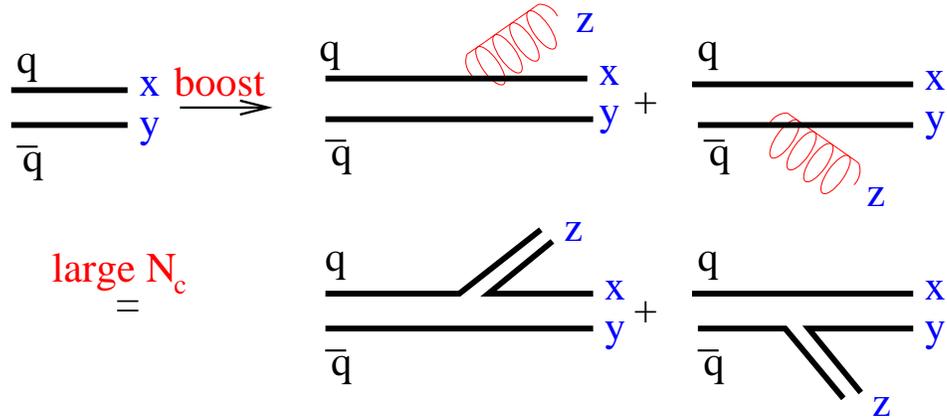,width=12.5cm}
\end{center}
\vskip -1.cm
\caption{Diagrams for gluon emission in the evolution of a dipole and its
$N_c\to \infty$ limit.}
\label{bkfig}
\end{figure}

The BK equation has the following probabilistic interpretation
\cite{Kovner:2000pt} (see Figure \ref{bkfig}):
when evolved in rapidity, the parent dipole with ends
located at $\vec{x}$ and $\vec{y}$ emits a gluon, which corresponds
in the large-$N_c$ limit to two dipoles with ends $(\vec{x},\vec{z})$ and
$(\vec{z},\vec{y})$, respectively. The probability of such emission is
given by the BFKL kernel (\ref{kernelfc}),
and weighted by the scattering probability of the
new dipoles minus the scattering probability of the parent dipole (as the
variation with rapidity of the latter is computed). The non-linear term is
subtracted in order to avoid double counting. It is this non-linear term which
prevents, in contrast to BFKL, the amplitude from growing boundlessly with
rapidity. The BK equation ensures unitarity locally in transverse
configuration space, $|N(r,Y)|\leq 1$. This is guaranteed since
for $N(r,Y)=1$, the derivative with respect to $Y$ in (\ref{eq:bk})
cannot be positive.

\section{Running coupling} \label{running}

The BK equation (\ref{eq:bk}) was
derived at leading order in $\alpha_s\ln{(s/s_0)}$ for a fixed coupling
constant $\alpha_s$. An important part of the next-to-leading-log
corrections is expected to come, as in BFKL, from the running of the coupling.
The scale of the running coupling can only be determined when the
next-to-leading-log calculation is available. In this paper we introduce
heuristically the running of the coupling, as done previously in BFKL (see
e.g. \cite{Lipatov:1985uk,Collins:1988ze}); we will use different
prescriptions for the
scales in order to check the sensitivity of the results. To motivate our
choices, we recall the interpretation of the BFKL kernel (\ref{kernelfc}) 
as the Weizs\"acker-Williams probability for gluon emission written in a 
dipolar form,
\beq
K(\vec{r},\vec{r}_1,\vec{r}_2)\equiv K_0
(\vec{r},\vec{r}_1,\vec{r}_2)=\frac{\alpha_sN_c}
{\pi}\,\frac{r^2}{r_1^2r_2^2}=\frac{N_c}{4\pi^2}\left\vert\frac{g_s\vec{r}_1}
{r_1^2}-\frac{g_s\vec{r}_2}{r_2^2}\right\vert^2,
\label{eq:sepker}
\eeq
with $g_s=\sqrt{4 \pi \alpha_s}$.

Three distance scales appear in this kernel: an `external'
one, the size of the parent dipole, $r$, and two `internal' ones,
the sizes of the two newly created dipoles, $r_1$ and $r_2$.
The latter depend on the transverse position of the emitted
gluon $\vec{z}$ and on $\vec{r}$ through (\ref{coord}). We study 
three different
prescriptions for implementing these scales in a running coupling
constant in the BFKL kernel (\ref{eq:sepker}):

\begin{enumerate}
\item
In the first modified kernel, $K1$, the scale at which the running 
of the coupling is evaluated is taken to be that of the size of the
parent dipole, $r$. This choice amounts to the
substitution $\alpha_s\rightarrow\alpha_s(r)$ in Equation
(\ref{eq:sepker}),
\beq
K_1(\vec{r},\vec{r}_{1},\vec{r}_{2})=\frac{\alpha_s(r)N_c}{\pi}\,
\frac{r^2}{r_1^2r_2^2}\,.
\label{eq:k1}
\eeq
\item
To implement the running of the coupling at the internal scale, we
alternatively modify the emission amplitude in (\ref{eq:sepker}) before 
squaring it,
\beq
K_2(\vec{r},\vec{r}_1,\vec{r}_2)=\frac{N_c}{4\pi^2}
\left|\frac{g_s(r_1)\vec{r}_1}{r_1^2}-\frac{g_s(r_2)\vec{r}_2}
{r_2^2}\right|^2.
\label{eq:k2}
\eeq
\item
In order to check the sensitivity of the results to the Coulomb tails of the
kernel, we further modify the kernel $K2$ by imposing short range 
interactions, so that the emission of large size dipoles is 
suppressed. To do this, we weight the gluon emission vertex 
by exponential (Yukawa-like) terms,
\bea
K_3(\vec{r},\vec{r}_1,\vec{r}_2)=\frac{N_c}{4\pi^2}
\left|\frac{e^{-\mu r_1/2}g_s(r_1)\vec{r}_1}{r_1^2}-
\frac{e^{-\mu r_2/2}g_s(r_2)\vec{r}_2}{r_2^2}\right|^2,
\label{eq:k3}
\eea
with $\mu=\Lambda_{QCD}$.
\end{enumerate}

Let us anticipate that the different prescriptions $K1$,
$K2$ and $K3$ lead to very similar results for the evolution.
This can be 
traced back to the fact that all the geometrical dependence on $\vec{z}$
is integrated out so that only the $r$ dependence in the 
running of the coupling survives. Even the introduction of the
exponential damping has little effect, unless the range of the
interaction is chosen unphysically small (i.e. $\mu\gg \Lambda_{QCD}$).
However, the inclusion of a short range damping effect is 
known \cite{Kovner:2001bh,Golec-Biernat:2003ym}
to alter significantly the solution of the BK equation with impact
parameter dependence, which we do not consider in the present work.

For the qualitative properties of BK evolution studied
in this paper, the precise value and running of the coupling constant
is unimportant. To be specific, we use the standard one loop 
expression
\beq
\alpha_s(r)=\alpha_s(k=2/r)=\frac{12\pi}{\beta_0\ln{\left
(\frac{4}{r^2\Lambda_{QCD}^2}+\lambda\right)}}\,,
\label{eq:rc}
\eeq
where $\lambda$ is an infrared regulator and $\beta_0=11N_c-2N_f$
with $N_f=3$.  Both $\lambda$ and $\Lambda_{QCD}$ are 
determined from the conditions $\alpha_s(r=\infty)=\alpha_0$,
$\alpha_s(r=2/M_{Z^0})=0.118$, where $M_{Z^0}$ is the mass of 
the $Z^0$ boson. In our work, this choice is not motivated by
phenomenology, but by its use in related works e.g.
\cite{Iancu:2002tr,Triantafyllopoulos:2002nz} to which we want to compare.
From now on, when comparing fixed and running coupling results,
it will be understood that the value for the fixed coupling is 
the same as the one at which the running coupling is frozen, $\alpha_0$.

\section{Numerical method and initial conditions} \label{numerical}

To solve the integro-differential equation (\ref{eq:bk}), 
we employ a second-order Runge-Kutta method
with a step size $\Delta Y=0.1$. We discretize the variable
$|\vec{r}|$ into 1200 points equally separated in logarithmic space 
between $r_{min}= 10^{-12}$ and $r_{max}=10^2$. The numerical values of these
limits are dictated
by the initial conditions and $\Lambda_{QCD}$. Throughout this paper,
the units of $r$ will be GeV$^{-1}$, and those of $Q_s$ GeV.
The integrals in (\ref{eq:bk}) are performed with the Simpson method. 
Inside the grid a linear interpolation is used.
For points lying outside the grid with $r<r_{min}$ a power-law extrapolation
is used, while for points with $r>r_{max}$ the saturated 
value of the scattering probability is held constant, $N(r) \equiv 
N(r_{max}) = 1$.
While the initial conditions of $N(r)$
give negligible values for $r$ small but much
larger than $r_{min}$, the evolution leads to a gradual filling of values
close to $r_{min}$ with increasing rapidity, which would result eventually
in
numerical inaccuracies.
To solve this problem and push the evolution to very large rapidity, we rescale,
in the fixed coupling case, the
variable $r$ in the solutions at intermediate values
of $Y$ and use them as initial condition (a power-law extrapolation is used
for small values of $r$ in order to cover the $r$-range lost in the rescaling
procedure).
In this way, we are able to evolve initial conditions with $Q_s\sim 1$ GeV up
to $Y\sim 36$ for $\bar{\alpha}_s=0.4$ and up to $Y\sim 72$ for
$\bar{\alpha}_s=0.2$. In the running coupling case, the evolution is much 
slower and this rescaling is not needed to get to large rapidities.
The accuracy of our numerical solution for all $r$-values inside
the grid is better than 4\% up to the largest rapidities. It is much better
than 4\% in most of the $r$-region studied. We have checked this numerical
accuracy by varying the step size in $Y$, by comparing our results to those
of a fourth-order Runge-Kutta method, by varying the limits of the grid, 
by doubling the number of points used to discretize the function in the grid,
and by using different integration, extrapolation and interpolation methods.

We evolve three different
initial conditions starting from some fixed value of $x_0$ (in
practice one usually takes $x_0\sim 0.01$).
The first initial condition we refer to as GBW since it shows at fixed $x_0$
the same $r$-dependence as the
Golec-Biernat--W\"usthoff
model \cite{Golec-Biernat:1998js}:
\beq
N^{GBW}(r)=1-\exp{\left[-\frac{r^2Q_s^{\prime 2}}{4}\right]}\,.
\label{eq:gbw}
\eeq
However, in contrast to the GBW model \cite{Golec-Biernat:1998js}, our
$x$-dependence comes from BK evolution and we do not
impose a power-law parameterization of the $x$-dependence of $Q_s^{\prime }$.
Here and in the other initial conditions 
(\ref{eq:mv}), (\ref{eq:as}) below, we denote as $Q_s^{\prime }$ what is
usually called the saturation scale.
Our definition of the saturation scale $Q_s$ is somewhat
different, see Equation (\ref{defqs}) below,
but the relation between both scales is
straightforward e.g. in GBW, $Q_s^{\prime 2}=-4\ln{(1-\kappa)}\, Q_s^2$.
The second initial condition takes the form given by the
McLerran-Venugopalan model \cite{McLerran:1993ni,McLerran:1993ka} (MV):
\beq
N^{MV}(r)=1-\exp{\left[-\frac{r^2Q_s^{\prime 2}}{4}
\ln{\left(\frac{1}{r^2\Lambda_{QCD}^2}+e\right)}\right]}\, .
\label{eq:mv}
\eeq
These initial conditions have been used in previous works e.g.
\cite{Armesto:2001fa,Albacete:2003iq}.
For transverse momenta $k \sim 1/r \geq
{\cal O}(1\, {\rm GeV})$, the sensitivity to the infrared cut off $e$ is 
negligible. The amplitudes $N^{GBW}$ and $N^{MV}$ are similar for
momenta of order $Q_s^\prime$ but differ strongly in their high-$k$
behaviour. The corresponding unintegrated gluon distribution
$\phi(k) = \int \frac{d^2r}{2\pi r^2} e^{i \vec{r}\cdot \vec{k}}\, N(r)$
decays exponentially for $N^{GBW}$ but has a power-law tail
$\sim 1/k^2$ for $N^{MV}$. 
As a third initial condition, we consider
\beq
N^{AS}(r)=1-\exp{\left[-(r\,Q_s^\prime)^c\right]}\, .
\label{eq:as}
\eeq
The interest in this ansatz is that the small-$r$ behaviour
$N^{AS} \propto r^c$ corresponds to an anomalous
dimension $1-\gamma = 1- c/2$ of the unintegrated gluon distribution
at large transverse momentum. This anomalous dimension can be chosen
to differ significantly from that of the initial conditions $N^{GBW}$ 
and $N^{MV}$. Our choices $c=1.17$ and $c=0.84$ are somewhat arbitrary.
They can be motivated a posteriori by the observation that 
the anomalous dimension of the evolved BK solution 
for both fixed and running coupling 
lies between the anomalous dimension of the initial conditions 
$N^{AS}$ and $N^{GBW}$ (or $N^{MV}$). 
Thus, the choice of $N^{AS}$ is very convenient to
establish generic properties of the solution of
the  BK equation.
The values of $Q_s^\prime$
in Equations (\ref{eq:gbw}), (\ref{eq:mv}) and
(\ref{eq:as}) are 1.4 GeV for GBW, 4.6 GeV for MV, 0.7 GeV for AS with
$c=1.17$ and 0.6 GeV for AS with $c=0.84$.
These values
have been used in all our
studies except in those on the $A$-dependence in Section
\ref{adep}, where $Q_s^\prime$
has been rescaled with the nuclear size as discussed
in that Section.

\section{Results} \label{results}

In this Section, we discuss our
numerical results and how they compare to
previous numerical work and
analytical estimates.

\subsection{Evolution:
Insensitivity to details of running coupling prescription} 
\label{evolution}

\begin{figure}[t]
\begin{center}
\epsfig{file=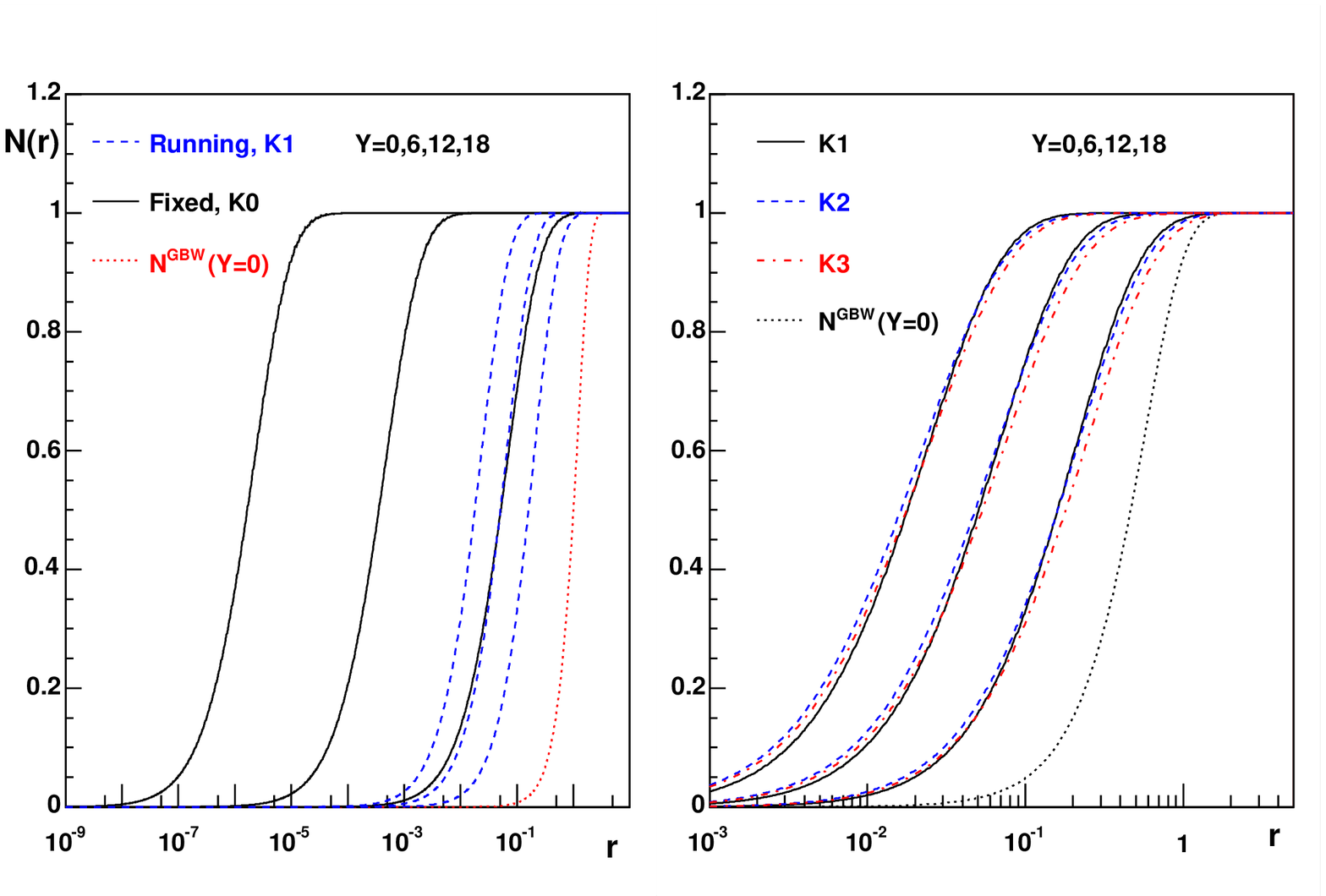,width=13.5cm}
\end{center}
\vskip -1.cm
\caption{Solutions of the BK equation for GBW initial condition (dotted line)
for rapidities $Y=6$, 12 and 18 with $\bar \alpha_0=0.4$.  Left plot:
Evolution with fixed ($K0$, solid lines) and running coupling ($K1$, dashed
lines). Right plot: evolution with running coupling for kernel modifications
$K1$ (solid lines), $K2$ (dashed lines) and $K3$ (dashed-dotted lines).}
\label{fig1}
\end{figure}

Figure \ref{fig1} shows the evolution of the dipole scattering 
probability for GBW initial condition with fixed and running 
coupling. The evolution is much faster for fixed coupling than 
for running coupling, as already known from previous numerical studies
\cite{Golec-Biernat:2001if,Braun:2003un,Rummukainen:2003ns}.
Remarkably, the solution is rather insensitive to the precise
prescription with which running coupling effects are implemented
in the modified BFKL kernels $K1$, $K2$ and $K3$. These
differences are very small compared to those
between fixed and running coupling. 

The small differences arising from the use of different kernels can 
be understood qualitatively. For example, compared to $K1$, the 
results obtained for $K2$ are enhanced at small
values of $r$ and suppressed at large values of $r$.
This is due to the fact that e.g. for a typical size $\sim 1/Q_s$ 
of the emitted dipoles $r_1$, $r_2$,
a larger size $r>1/Q_s$ of the parent
dipole amounts to a larger coupling $g_s(r)$ entering the kernel $K1$
than the couplings $g_s(r_1)$, $g_s(r_2)$ entering $K2$. Thus, at large
$r$ the evolution is slower for $K2$, which results in the observed 
relative suppression. The
analogous argument implies a relative enhancement obtained from the
kernel $K2$ for small $r<1/Q_s$.

Figure \ref{fig1} also shows that the effects of imposing 
short range interactions, $K3$, are very small (unless the range of 
the interaction is unphysically small). As expected, effects from
short range interactions included in $K3$ are larger for larger values 
of $r$. It is conceivable that the main next-to-leading-log effects on the 
original BK kernel are those of the running of the coupling constant
included here, and that further modifications, such us kinematical 
constraints
\cite{Schmidt:1999mz,Forshaw:1999xm,Golec-Biernat:2001if},
are comparatively small \cite{Chachamis:2004ab}. 

\subsection{Scaling} \label{scaling}

In the limit $Y \to \infty$, the solutions of the BK evolution are no 
longer functions of the variables $r$ and $Y$ separately, but instead they 
depend on a single scaling variable
\beq
\tau\equiv r\,Q_s(Y)\, .
\eeq
Here, the saturation momentum $Q_s(Y)$ determines the transverse
momentum below which the unintegrated gluon distribution is saturated.
It can be characterized
by the position of the falloff in $N(r)$,
e.g. via the 
definition
\beq
  N(r=1/Q_s(Y),Y)=\kappa,
\label{defqs}
\eeq
where $\kappa$ is a constant which is smaller than, but of order, one.
We have checked that different choices such as $\kappa=1/2$ and $\kappa=1/e$
lead to negligible differences in the determination of $Q_s(Y)$. The
results given below have been obtained for $\kappa=1/2$.

\begin{figure}[t]
\begin{center}
\epsfig{file=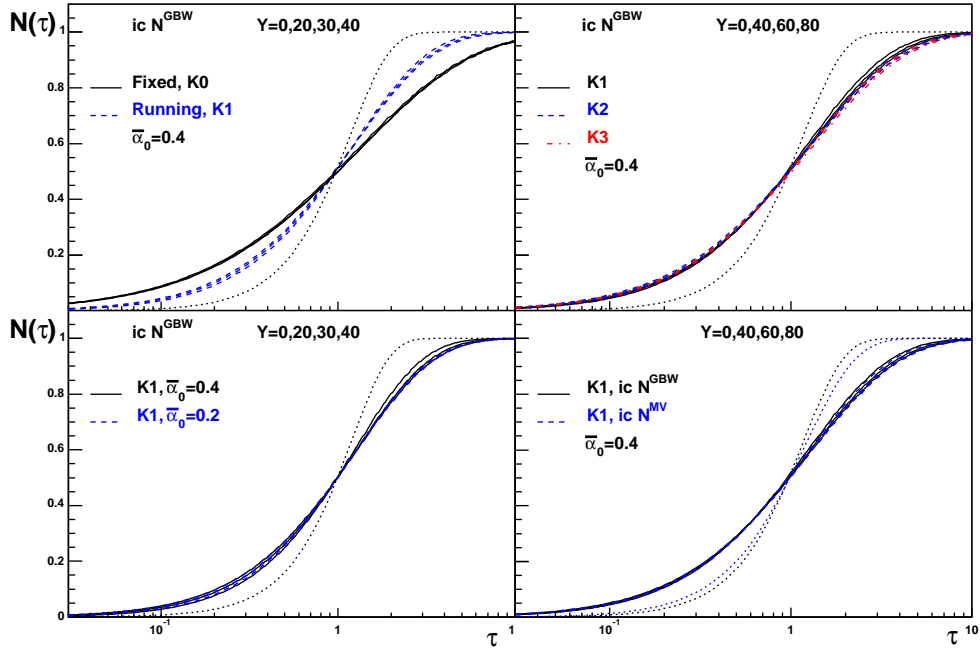,width=13.5cm}
\end{center}
\vskip -1.cm
\caption{Scaling solutions of BK for $Y=0$, 20, 30 and 40 (plots on the left)
and $Y=0$, 40, 60 and 80 (plots on the right). Upper-left: evolution for fixed
(solid) and running coupling ($K1$, dashed lines) for GBW initial conditions.
Upper-right: solutions for the kernels $K1$
(solid), $K2$ (dashed) and $K3$ (dashed-dotted lines). Lower-left: scaling
function for $K1$ with two different values of frozen coupling, $\bar
\alpha_0=0.4$ (solid) and $\bar \alpha_0=0.2$ (dashed lines). Lower-right:
scaling solutions with running coupling ($K1$) for two different initial
conditions, GBW (solid) and MV (dashed lines). In all plots the initial
conditions correspond to the dotted lines and $\bar \alpha_0=0.4$ unless
otherwise stated.}
\label{fig2a}
\end{figure}

In the fixed coupling case, the scaling property $N(r,Y) \to N(\tau)$
has been quantified in previous numerical works
\cite{Armesto:2001fa,Lublinsky:2001bc,Albacete:2003iq} and confirmed by
analytical calculations \cite{Munier:2003vc,Munier:2003sj,Munier:2004xu}.
In the running coupling case, the scale invariance of the BFKL kernel is 
broken by the scale $\Lambda_{QCD}$ and it is a priori unclear whether
scaling persists. However, when the two scales in the problem are 
separated widely due to evolution to large rapidity, 
$Q_s(Y)\gg \Lambda_{QCD}$, one may expect that the scaling property of 
the BK solution is restored. In agreement with previous numerical works
\cite{Braun:2003un,Rummukainen:2003ns}, we confirm this expectation:
for all modifications $K1$, $K2$ and $K3$ of the BFKL kernel, the solutions 
tend to universal scaling forms as rapidity increases. Moreover, with
increasing rapidity the sensitivity to the choice of scales in the kernel and
its short range modification, as well as
to the initial condition
and to the value of the
coupling constant in the infrared, becomes eventually negligible
(see Figure \ref{fig2a}).

As seen in Figure \ref{fig2a},
the shape of the scaling solution differs 
significantly for fixed and running coupling as observed already 
in \cite{Braun:2003un}. The running of the coupling 
suppresses the emission of dipoles of small transverse size 
(i.e. small $\tau$ and large transverse momenta). This leads to
an enhancement in the large $\tau$ region of $N(\tau)$ which is seen for the 
running coupling case in Figure \ref{fig2a}.

The accuracy of scaling at small $r$
has been studied in a previous work \cite{Albacete:2003iq} for the fixed
coupling case. Here we check
scaling for both fixed and running coupling
by comparing our numerical results to the scaling forms proposed in
Reference \cite{Mueller:2002zm}. There, it was argued
that in the so-called scaling window $\tau_{\rm sw}<\tau <1$, the 
asymptotic solution of $N(r,Y)$ takes the following scaling forms
for fixed and running coupling, respectively \cite{Mueller:2002zm}:
\beq
f^{1)}(\tau)=a\tau^{2\gamma}\left(\ln{\tau^2}+\delta\right)\, ,
\label{eq:f1}
\eeq
\beq
f^{2)}(\tau)=a\tau^{2\gamma}\left(\ln{\tau^2}+{1\over \gamma}\right)\, .
\label{eq:f2}
\eeq
Here, $1-\gamma$ is 
usually called the anomalous dimension which governs the leading
large-$k$ behaviour of the unintegrated gluon distribution.
We define $\gamma$ from a fit of our numerical results to the functions
(\ref{eq:f1}) and
(\ref{eq:f2}) in the $Y$-independent region
$10^{-5}<\tau<10^{-1}$, i.e. for $10^5\,Q_s>1/r>10\,Q_s$,
with $a$, $\gamma$ and $\delta$ as free parameters.
The results given below were found to be insensitive to a variation of
the lower limit of this fitting range.

\begin{figure}[t!]
\begin{center}
\epsfig{file=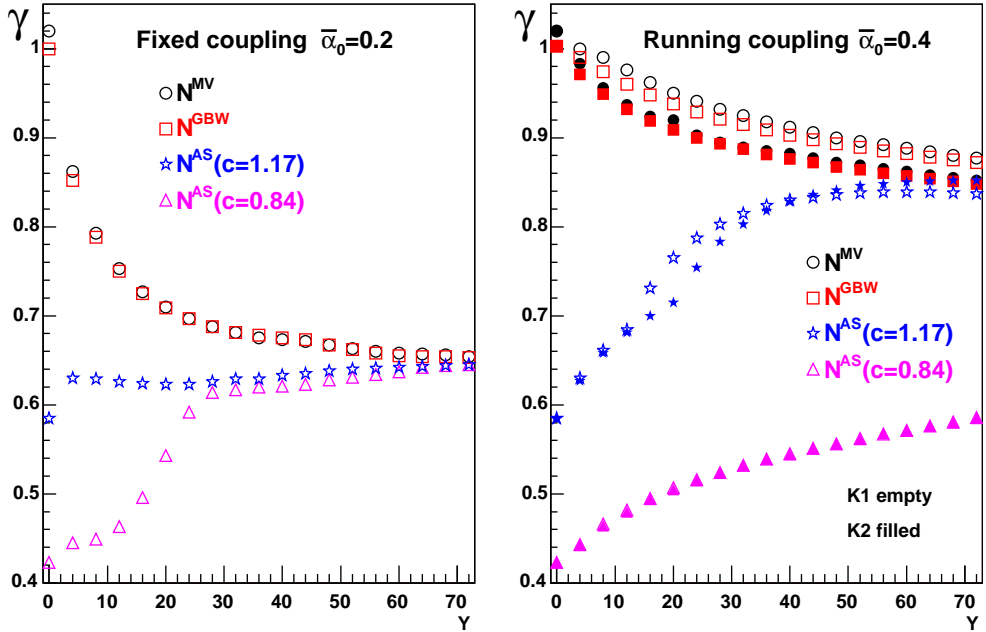,width=13.5cm}
\end{center}
\vskip -1.cm
\caption{The rapidity dependence of the parameter $\gamma$, characterizing
the anomalous dimension $1-\gamma$, as determined by a fit of  
(\ref{eq:f1}) to the BK
solutions for different initial conditions: GBW (squares),
MV (circles), and AS with $c=1.17$ (stars) and $c=0.84$ (triangles).  Left
plot: results for fixed coupling with $\bar \alpha_0=0.2$. Right plot: results
for running coupling with $\bar \alpha_0=0.4$ and two versions of the kernel
$K1$ (empty symbols) and $K2$ (filled symbols).}
\label{fig2b}
\end{figure}

For the case of fixed coupling constant, we find that the function
$f^{1)}$ provides a very good fit to the evolved solutions.
In Figure \ref{fig2b}, we show the fit values of the parameter
$\gamma$, obtained for fixed coupling constant from the
evolution of different initial conditions $N^{GBW}$, $N^{MV}$, and $N^{AS}$
for different values of $c$.
At initial rapidity, these distributions have widely different anomalous
dimensions but evolution drives them to a common
value, $\gamma\simeq 0.65$, which lies close to the theoretically 
conjectured one \cite{Iancu:2002tr,Mueller:2002zm} of $0.628$.
For a small fixed coupling constant $\bar{\alpha}_0=0.2$, this asymptotic
behaviour is reached at $Y \sim 70$, while for a larger coupling constant
$\bar{\alpha}_0=0.4$ the approach to this asymptotic value takes 
half the length of evolution (results not shown). For
fixed coupling solutions, $f^{2)}$ does not provide a good fit to our
numerical results.

We have repeated this comparison for all running coupling solutions.
We found that both $f^{1)}$ and $f^{2)}$ provide good fits and yield
very similar values of $\gamma$. The results for $K3$ are numerically
indistinguishable from those for $K2$ and will not be shown in what
follows. Also, the value of $\gamma$ was found to be independent
of the coupling constant $\bar \alpha_0$ at $r\to \infty$.
In Figure \ref{fig2b}, we show the values of $\gamma$ extracted 
from a fit to $f^{1)}$. Irrespective of the initial condition, they
approach a common asymptotic value $\gamma\sim 0.85$. While our
numerical findings for $N^{AS}$ with $c=0.84$ are not inconsistent
with the approach to this asymptotic value, no firm conclusions can
be drawn. This initial condition just starts too far
away from the asymptotic scaling solution to reach it within the 
numerically accessible rapidity range. In this case, the monotonic 
increase of $\gamma$ with rapidity at large $Y$ is smaller than the 
increase for $N^{AS}$ with $c=1.17$ at comparable values of $\gamma$,
indicating that the rapidity evolution of the anomalous dimension
depends in general not only on the small-$r$ behaviour, but
on the full shape of the scattering probability. 
 
The value $\gamma\sim 0.85$ is considerably larger
than the one found in fixed coupling evolution. This is
in agreement with previous numerical results \cite{Braun:2003un} but
in contrast to theoretical 
expectations \cite{Iancu:2002tr,Mueller:2002zm,Triantafyllopoulos:2002nz} 
which predict the same
value of $\gamma$ for the fixed and running coupling cases.
As an additional check, we have performed running coupling evolution from an
initial condition given by the solution at large
rapidity of fixed coupling evolution
(for which $\gamma \simeq 0.65$). We find that even with this
initial condition,
running coupling
evolution leads to a value of $\gamma \sim 0.85$.

\begin{figure}[t!]
\begin{center}
\epsfig{file=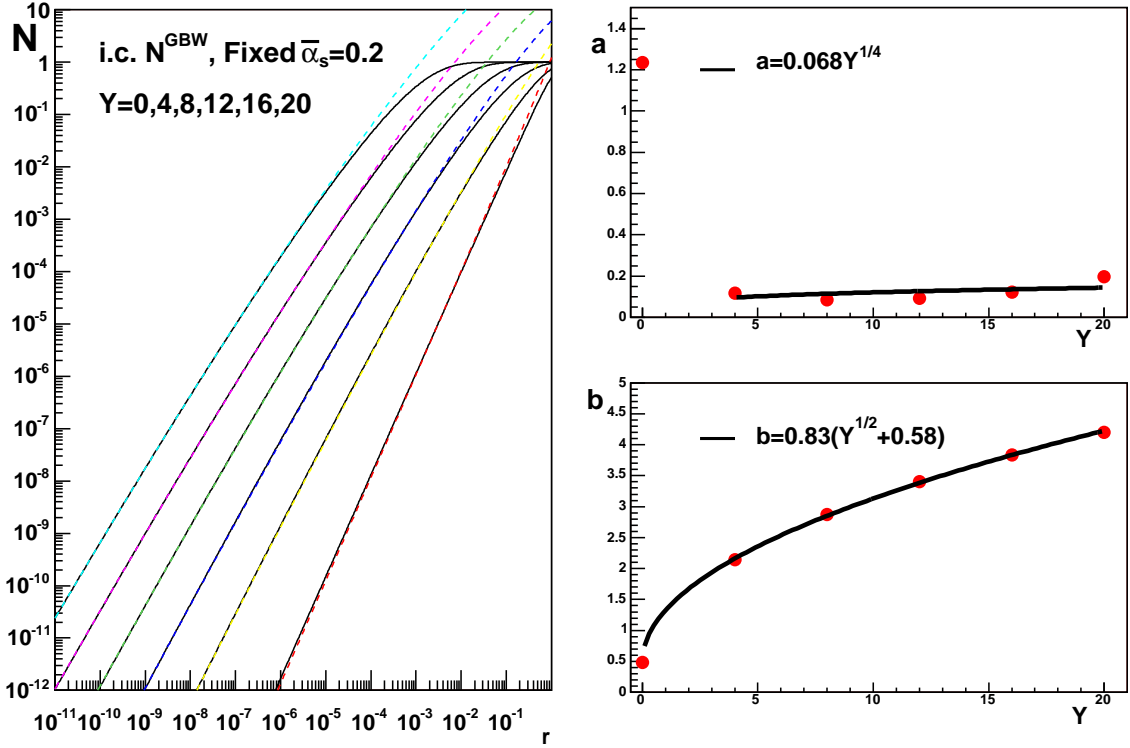,width=15.5cm}
\end{center}
\vskip -1.cm
\caption{Plot on the left: solutions of the BK equation (solid lines) with GBW
initial condition and fixed coupling $\bar \alpha_s=0.2$ compared to fits
(dashed lines) to the DLL expression (\ref{eq:dll}),
for rapidities $Y=0$, 4, 8, 12, 16 and
20 (curves from right to left). Plots on the right: values of the coefficients
$a(Y)$ and $b(Y)$ (circles)
in the DLL expression versus $Y$, compared to fits (curves) to the
functional form suggested by DLL.}
\label{figdll}
\end{figure}

It has been
argued \cite{Iancu:2002tr,Mueller:2002zm} that expressions (\ref{eq:f1}) and
(\ref{eq:f2}) are only
valid for values of $\tau$ inside the scaling window,
$\tau_{\rm
sw}\sim \Lambda_{QCD}
/Q_s(Y)<\tau\lesssim  1$ with $Y_0$ the initial rapidity,
and that the dipole
scattering probability returns to the double-leading-log (DLL) expression
\begin{equation}
N^{DLL}(r)=a(Y)
\,r^2\,[-\ln{(r^2\Lambda^2)}]^{-3/4}\,\exp{\left[b(Y)\sqrt{-\ln{(r^2\Lambda^2)}}
\,\right]}\,,
\label{eq:dll}
\end{equation}
with $a(Y)\propto Y^{1/4}$ and $b(Y)\propto \sqrt{Y}$,
for values $\tau < \tau_{\rm sw}$.
We have checked that this form provides a good fit
(fit and numerical solution
differ by less
than $\pm 10$\%)
to the fixed coupling
solution of BK
for $\tau<\tau_{\rm sw}=
\Lambda/Q_s(Y)$, $\Lambda\sim 0.2$ GeV,
see Figure \ref{figdll}.
Our comparison is limited to rapidities $Y\leq 20$, since the scaling window
starts to extend over the entire numerically accessible $r$-space for $Y>20$.
Up to $Y=20$, the coefficients $a(Y)$ and $b(Y)$ follow the expected
DLL $Y$-behaviour,
see Figure \ref{figdll}.
However, the
scaling ansatz $f^{1)}$ provides an equally good fit to the BK solutions 
for $\tau<\tau_{\rm sw}$.
This is the reason why in previous numerical studies  
\cite{Albacete:2003iq} no upper bound for a scaling window was found.
When the solutions of BK are fitted to $f^{1)}$
within the scaling window,
the values of $\gamma$ at $Y=0$ for both initial conditions are $\lesssim 20$\%
smaller than those found when the fit is done
within a fixed $\tau$-window.
But for larger $Y$
the values of $\gamma$ extracted from fits within either the
scaling window or some fixed $\tau$-window approach each other and quickly
coincide.

\subsection{Rapidity dependence of the saturation scale} \label{ydep}

In the scaling region, for large $Y$ where $Q_s(Y)\gg \Lambda_{QCD}$,
the BK Equation (\ref{eq:bk}) for fixed coupling constant can be
written in terms of the rescaled variables
$\vec{\tau}=Q_s(Y)\vec{r}$, $\vec{\tau}_1=Q_s(Y)\vec{r}_1$ and
$\vec{\tau}_2=Q_s(Y)\vec{r}_2$. The $Y$-dependence of
$N(r,Y)\equiv N(\tau)$ is then contained in $Q_s(Y)$. 
Rewriting the derivative on the left-hand-side of (\ref{eq:bk}),
\beq
{\partial N(\tau)\over \partial Y}={\partial Q_s(Y)\over \partial Y}\,r\,
{\partial N\over \partial \tau}={\partial \ln{[Q_s^2(Y)/\Lambda^2]}\over
\partial Y}\,r^2 \, {\partial N\over \partial r^2}\, ,
\label{reas1}
\eeq
one finds \cite{Iancu:2002tr}
\beq
\int {d^2r\over r^2} {\partial N(\tau)\over \partial Y} =
\pi\,{\partial \ln{[Q_s^2(Y)/\Lambda^2]}\over
\partial Y} \,[N(\infty)-N(0)]=\pi\,{\partial \ln{[Q_s^2(Y)/\Lambda^2]}\over
\partial Y}\,.
\label{reas2}
\eeq
Performing the same integration over $d^2r/r^2=d^2\tau/\tau^2$ on the
right-hand-side of (\ref{eq:bk}), one finds a number 
\beq
d=\int{d^2\tau d^2\tau_1\over 2 \pi^2}\,{1\over \tau_1^2
\tau_2^2}\, [N(\tau_1)+N(\tau_2)-N(\tau)-N(\tau_1)N(\tau_2)],
\label{reas3}
\eeq
which is independent of $Y$.
The numerical value of $d$ cannot be obtained without the knowledge of the
scaling solution $N(\tau)$, and several approximations have been proposed
\cite{Iancu:2002tr,Mueller:2002zm} which we will compare with our numerical
results. Combining Equations (\ref{eq:bk}), (\ref{reas2}) and (\ref{reas3}),
the $Y$-dependence of the saturation scale is
determined \cite{Iancu:2002tr} by
\beq
\frac{\partial\ln{\left[Q^2_s(Y)/\Lambda^2\right]}}{\partial Y}=d\,\bar
\alpha_s\, .
\label{qsy}
\eeq
Thus, for the case of a fixed coupling constant, the saturation scale
grows exponentially with rapidity,
\beq
Q_s^2(Y)=Q_0^2\,\exp{\left[\Delta Y\right]},
\label{qsyfix}
\eeq
where $\bar \alpha_s=\bar \alpha_0=$ constant, $\Delta=d\bar \alpha_0$ and 
$Q^2_0=Q^2_s(Y=0)$ (i.e. the evolution starts at $Y=0$).

For running coupling, the momentum scale is expected to be $\sim Q_s(Y)$.
This suggests the substitution
$\bar \alpha_s\to \bar \alpha_s(Q_s(Y))$ in Equation (\ref{qsy}).
To see this explicitly, let us include, as in $K1$, the coupling constant
$\bar \alpha_s(r)$
in the integrand of (\ref{reas3}), which leads to
\bea
d\,\bar
\alpha_s&\longrightarrow&
{12 N_c\over \beta_0}
\int{d^2\tau d^2\tau_1\over 2 \pi^2}\,{1\over \tau_1^2
(\vec{\tau}-\vec{\tau}_1)^2}\, {1\over
\ln{[Q_s^2(Y)/\Lambda_{QCD}^2]}-\ln{(\tau^2/4)}}\nonumber \\
&\times&[N(\tau_1)+N(|\vec{\tau}-\vec{\tau}_1|)
-N(\tau)-N(\tau_1)N(|\vec{\tau}-\vec{\tau}_1|)].
\label{reas4}
\eea
For $\tau\gg 1$, the integrand vanishes. For $\tau\ll 1$, 
the integral in $d^2\tau_1$ is finite and the remaining $d^2\tau$ 
suppresses the contribution of small $\tau$. So we conclude that the 
dominant region is that of $\tau\sim 1$ and thus it is legitimate 
to approximate
\bea
{12 N_c\over \beta_0}
\int{d^2\tau d^2\tau_1\over 2 \pi^2}\,{1\over \tau_1^2
(\vec{\tau}-\vec{\tau}_1)^2}\, {1\over
\ln{[Q_s^2(Y)/\Lambda_{QCD}^2]}-\ln{(\tau^2/4)}}&&\nonumber \\
\times \ \ [N(\tau_1)+N(|\vec{\tau}-\vec{\tau}_1|)
-N(\tau)-N(\tau_1)N(|\vec{\tau}-\vec{\tau}_1|)]&\simeq& d\,
\bar \alpha_s(Q_s(Y)).
\label{reas5}
\eea
This approximation is also supported by numerical results 
\cite{Braun:2000wr,Armesto:2001fa,Golec-Biernat:2001if} which show
that in momentum space the typical transverse momentum of the gluons 
is $\sim Q_s$. Due to the similarities in the evolution shown
previously, this should also hold for other implementation of 
the scale of the coupling constant such as $K2$ and $K3$.
The logarithmic dependence of the coupling constant on $Q_s(Y)$ 
in (\ref{reas5}), combined with Equation (\ref{qsy}), leads to
\cite{Iancu:2002tr}
\beq
Q_s^2(Y)=\Lambda^2 \,\exp{\left[\Delta^\prime\sqrt{Y+X}\right]},
\label{qsyrun}
\eeq
where $(\Delta^\prime)^2=24N_c d/\beta_0$
and $X=(\Delta^\prime)^{-2}\ln{(Q_0^2/\Lambda^2)}$. This estimate
indicates that the rapidity dependence 
of the saturation scale is much weaker for running than for fixed coupling
constant.

\begin{figure}[t!]
\begin{center}
\epsfig{file=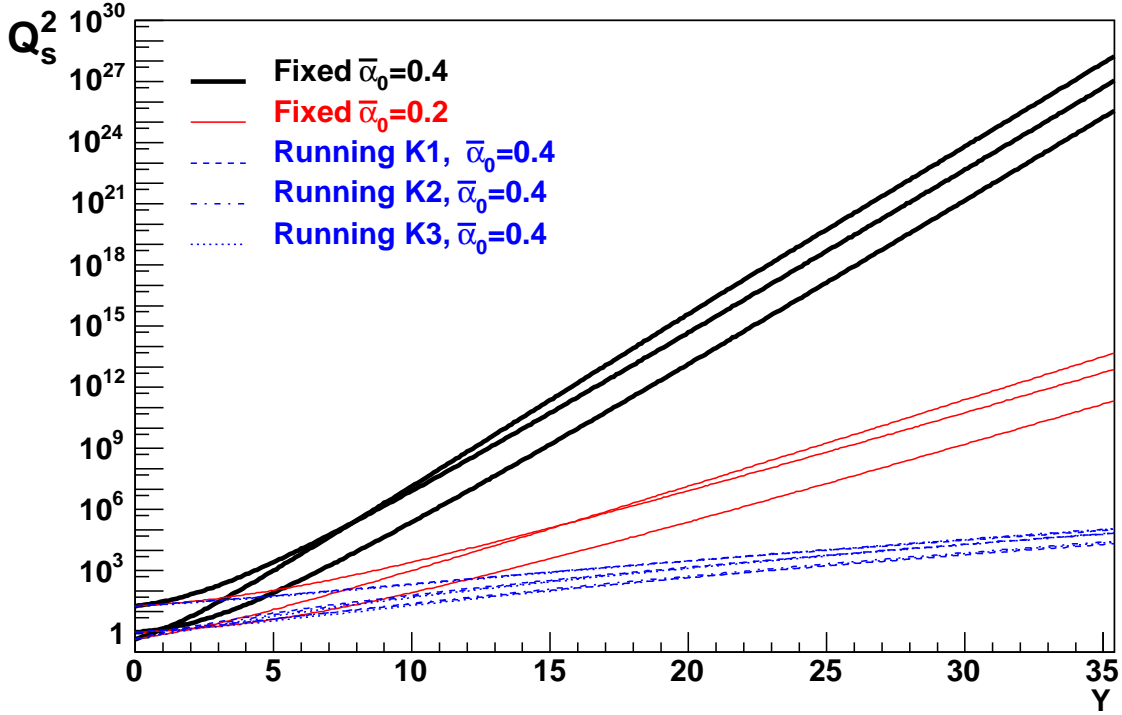,width=15.5cm}
\end{center}
\vskip -1.cm
\caption{The rapidity dependence of the saturation momentum $Q_s^2$ 
for fixed $\bar \alpha_s=0.4$ (thick solid),
fixed $\bar \alpha_s=0.2$ (thin solid), and running coupling with $\bar
\alpha_0=0.4$ for kernels $K1$ (dashed), $K2$ (dashed-dotted) and $K3$ (dotted
lines).  For each group, lines from top to bottom in the rightmost side
correspond to initial conditions AS with $c=1.17$, MV and GBW.}
\label{fig3}
\end{figure}

Figure \ref{fig3} shows the $Y$-dependence of $Q_s^2$ for several 
initial conditions and different choices of $\bar \alpha_0$, calculated
for all the 
kernels considered in this work. The rise of $Q_s$ is much faster 
for fixed than for running coupling, as already observed in 
\cite{Lublinsky:2001yi,
Golec-Biernat:2001if,
Triantafyllopoulos:2002nz,Braun:2003un,Rummukainen:2003ns,Kutak:2004ym,
Khoze:2004hx,Chachamis:2004ab}.

For fixed coupling constant, $Q_s^2$ exhibits with good accuracy
an exponential
behaviour for high-enough values of $Y$. The value of the slope 
extracted from a fit to the function (\ref{qsyfix}) is $\Delta\simeq 1.83$
for $\bar \alpha_0=0.4$. As expected, for $\bar \alpha_0=0.2$ this value 
is reduced by a factor two, $\Delta\simeq 0.91$. For the constant 
(\ref{reas3}), we find $d\simeq 4.57$, in agreement
with previous numerical studies at very high rapidities \cite{Albacete:2003iq}
but slightly smaller than the theoretical expectation $d=4.88$
\cite{Iancu:2002tr,Mueller:2002zm}.
In previous numerical studies 
\cite{Armesto:2001fa,Levin:2001et,Golec-Biernat:2001if},
an even smaller value of $d \sim 4.1$ was obtained. We have checked that 
this is due to the fact that the rapidity region for the fit in our case 
corresponds to much larger $Y$.

For the case of a running coupling constant, an exponential fit 
can be done only for a very limited $Y$-region. For example, for 
$Y\sim 10$ we find a logarithmic slope $\sim0.28$ for GBW or MV initial
conditions with $Q_0 \sim 1$ GeV, in
agreement with the results of \cite{Triantafyllopoulos:2002nz} but
smaller than the values found in
\cite{Khoze:2004hx} (see also
\cite{Kutak:2004ym,Chachamis:2004ab}).
The exponential function (\ref{qsyfix}) is unable to fit the full
$Y$-range. In contrast, the weaker rapidity dependence of (\ref{qsyrun}) 
does provide a good fit in the full $Y$-range.
The fit to (\ref{qsyrun})  
yields $\Delta^\prime \simeq 3.2$, while the theoretical expectation
\cite{Iancu:2002tr,Mueller:2002zm} is slightly larger, $\Delta^\prime=3.6$.
We finally note that in \cite{Rummukainen:2003ns} the $Y$-derivative of 
$\ln{Q_s^2(Y)}$ has been
found numerically to be
proportional to $\sqrt{\alpha_s(Q_s(Y))}$ in a much more
restricted range of $Y$. We have been unable to fit our results over 
the full $Y$-range to the corresponding $Y$-dependence, $Q_s^2(Y)\propto
\exp{Y^{2/3}}$.

We have found very little
sensitivity of the values of $\Delta$ and $\Delta^\prime$ 
to the fitting region, provided $Y$ was chosen large enough. Our fits typically
started at $Y\sim 15$ where the asymptotic behaviour is approached,
and explored the highest rapidities numerically accessible.
Also, our results for running coupling do not depend on the choice of
the kernel $K1$, $K2$ or $K3$, on the
initial condition or on the value of $\bar \alpha_0$.
However, the AS initial condition with $c=0.84$
is not included in our study since it does not approach the asymptotics
within the numerically accessible rapidity range.

In References
\cite{Mueller:2002zm,Munier:2003vc,Munier:2003sj,Munier:2004xu,
Triantafyllopoulos:2002nz}
sub-leading terms in the $Y$-behaviour of $Q_s$ have been presented. A
form of the type $d\ln{Q_s^2(Y)}/ dY=\bar \alpha_s a -b
Y^{-1}+cY^{-3/2}/(2\sqrt{\bar \alpha_s}\,)$
has been proposed in the fixed coupling case, with $a=4.88$,
$b=2.39$ and $c=2.74$.
This function contains all terms for the $Y$-evolution of the
saturation scale that are universal i.e.
independent of the initial condition
(see also
\cite{Golec-Biernat:2004sx} for a comparison
of solutions of BK to this functional
form). The constant term corresponds to Equation
(\ref{qsyfix}). We have used this
functional form to fit the results of fixed coupling evolution
on $d\ln{Q_s^2(Y)}/ dY$ for different
rapidity regions within $Y=5\div 40$ (72) for $\bar \alpha_s=0.4$ (0.2), for
the GBW and MV initial conditions.
First, we have used our  definition of the saturation scale (\ref{defqs}) with
$\kappa=1/2$.
From a simple comparison to
the proposed expression (using the theoretical coefficients provided in
\cite{Munier:2004xu}), we are able to clearly identify in
our numerical results the presence of the
first two terms.
On the contrary, the presence of the third term is
disfavoured.
Fitting our numerical results to the first plus
second terms, the value of $a$ we
find, $a\simeq 4.9$, is quite stable with respect to variations of the fitting
region. It is higher than the value of $d$ we extract with only the linear
term (\ref{qsyfix}),
$d\simeq 4.57$, and closer to the theoretical expectation $d=4.88$
\cite{Iancu:2002tr,Mueller:2002zm}.
In this two-parameter fit
we get a value of $b\simeq 2.4\div 2.5$, varying slightly with
the $Y$-region of the fit. This value is quite close to the theoretical
expectation 2.39. On the other hand,
in a three-parameter fit the values of $b$ and $c$
we extract are very unstable (even changing signs) with respect to variations
of the lower limit of the fitting region between $Y=5$ and 20.
We have also tried to get the value of $c$ from
a fit to $d/dY[Y\,d\ln{Q_s^2(Y)}/ dY]=\bar \alpha_s a
-cY^{-3/2}/(4\sqrt{\bar \alpha_s})$. While we find again
a value of $a\simeq 4.9$, the value of $c$ turns out to
depend, as in the previous analysis, considerably on
the fitted $Y$-region.
Secondly,
we have used the definition of the saturation scale (\ref{defqs}) but now
with
$\kappa=0.01$ (i.e. we define $Q_s$
in a point in which the dipole scattering probability is
far from its unitarity limit). In this case, a simple comparison to
the proposed expression using the theoretical coefficients provided in
\cite{Munier:2004xu} allows to clearly identify in
our numerical results the presence of the
three terms. Still, a three-parameter fit to our numerical results
does not provide values of $b$ and
$c$ stable with respect to changes in the fitting region. This influence of the
definition of the saturation scale on the determination of the sub-leading
corrections to its $Y$-behaviour
is consistent with the finding in
\cite{Golec-Biernat:2004sx}.

\subsection{Nuclear size dependence of the saturation scale} \label{adep}

The nuclear size enters the initial condition. The question 
is whether the BK evolution modifies or preserves this initial $A$-dependence. 
For realistic nuclei, the impact parameter is likely to have 
an important effect on this $A$-dependence. This has 
been examined partially in \cite{Golec-Biernat:2003ym,Gotsman:2004ra}. 
However, the question is already of interest for the case
without impact parameter dependence 
\cite{Armesto:2001fa,Levin:2001et,Mueller:2003bz}, which we study here.

Let us first assume some arbitrary $A$-dependence which we include
in the initial condition by the rescaling factor
\beq
r^2 \longrightarrow h\, r^2
\label{nucres}
\eeq
(this is true for GBW and AS initial conditions but not for MV due to the
presence of the logarithm; however,
the numerical results for the $A$-dependence obtained
with MV initial conditions are, for all purposes,
equivalent to those with GBW).
Here, $h$ contains the information about the nuclear size, 
and Equation (\ref{qsy}) reads
\beq
\frac{\partial\ln{\left[Q^2_s(Y)/\Lambda^2\right]}}{\partial Y}=d\,\bar
\alpha_s(\sqrt{h}\,Q_s(Y))\, .
\label{qsyp}
\eeq
In the case of a fixed coupling constant, the
dilatation invariance of the BK equation (\ref{eq:bk}) allows
to scale out any nuclear dependence included in the initial condition. Thus,
the $A$-dependence of the saturation scale is unaffected by evolution.
To explore the case of a running coupling constant, we use 
the one-loop expression for $\alpha_s$ and write 
\beq
Q_s^2(Y)={\Lambda^2\over h}\, \exp{\sqrt{(\Delta^\prime)^2Y+\ln^2{\left[
{h\,Q_s^2(Y=0)\over \Lambda^2}\right]}}}\, .
\label{qsya}
\eeq
Multiplying by $h$ for the nucleus to undo the rescaling,
and setting $h=1$ for the proton, we get
\beq
{Q_{sA}^2(Y)\over Q_{sp}^2(Y)}=
\exp{\left\{\sqrt{(\Delta^\prime)^2Y+\ln^2{\left[
{h\,Q_s^2(Y=0)\over \Lambda^2}\right]}}-\sqrt{(\Delta^\prime)^2Y+\ln^2{\left[
{Q_s^2(Y=0)\over \Lambda^2}\right]}}\right\}}\,.
\label{ratqsya}
\eeq
If we assume the hierarchy
\beq
(\Delta^\prime)^2Y\gg \ln^2{\left[
{h\,Q_s^2(Y=0)\over \Lambda^2}\right]}
\gg \ln^2{\left[
{Q_s^2(Y=0)\over \Lambda^2}\right]}\,,
\label{hier}
\eeq
so $A\gg 1$, we find
\beq
\ln{{Q_{sA}^2(Y)\over Q_{sp}^2(Y)}}\simeq {\ln^2{\left[
{h\,Q_s^2(Y=0)\over \Lambda^2}\right]}\over 2\sqrt{(\Delta^\prime)^2Y}}\,.
\label{finmue}
\eeq
Here, $h\,Q_s^2(Y=0)$ is the initial saturation momentum for the nucleus, 
and Equation (\ref{finmue}) coincides with Equation (44) of 
\cite{Mueller:2003bz}
with $(\Delta^\prime)^2$ as defined below Equation (\ref{qsyrun})
(see also \cite{Rummukainen:2003ns,Kharzeev:2004if,Levin:1987xr} for
related discussions).
This result suggests that any information about the initial $A$-dependence
of the saturation scale is gradually lost during evolution: albeit
at extremely large rapidities, all hadronic targets look the same.
Usually, one assumes an $A^{1/3}$-dependence of the saturation scale 
for the initial condition \cite{McLerran:1993ni,McLerran:1993ka}, 
$h \propto A^{1/3}$. However,
other $A$-dependencies
have been proposed e.g. \cite{Levin:2001cv}.

\begin{figure}[t!]
\begin{center}
\epsfig{file=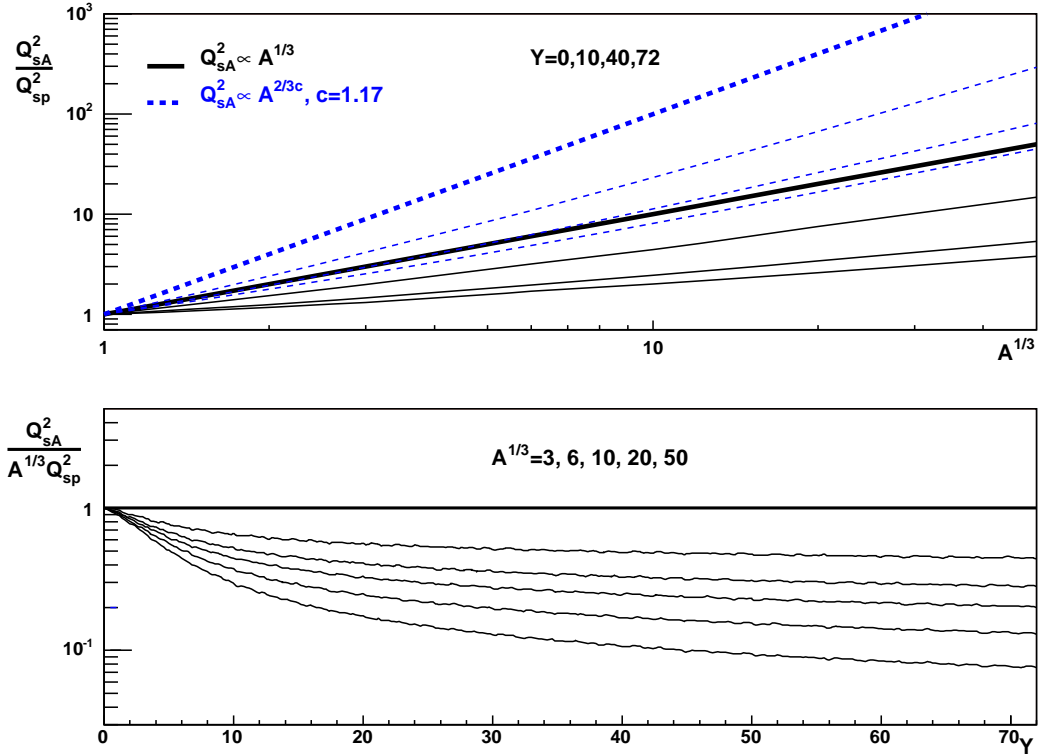,width=15.5cm}
\end{center}
\vskip -1.cm
\caption{Upper plot: $Q_{sA}^2/Q_{sp}^2$ versus $A^{1/3}$ for initial
conditions GBW ($Q_{sA}^2(Y=0)\propto A^{1/3}$, solid) and AS with $c=1.17$
($Q_{sA}^2(Y=0)\propto A^{2/3c}$, dashed lines); thick lines are the
results for $Y=0$ in the running coupling case and for all rapidities in fixed
coupling; for running coupling, different rapidities $Y=10$, 40 and 72 (thin
lines) are
shown from top to bottom for each initial condition. Lower plot:
$Q_{sA}^2/(A^{1/3}Q_{sp}^2)$ versus $Y$ for GBW with
$A^{1/3}=3$, 6, 10, 20 and 50 with
the same line convention as the upper plot (the results for fixed coupling
have been obtained for $Y<36$ and extrapolated as a constant equal to 1).
In all plots
$\bar \alpha_0=0.4$ and in the running coupling case the kernel $K1$ has been
used.}
\label{fig5}
\end{figure}

Figure \ref{fig5} shows that fixed coupling 
evolution preserves the $A$-dependence of the saturation scale
irrespective of whether this dependence is $\propto A^{1/3}$ 
as for the GBW or MV initial conditions (which produces numerical results for
the $A$-dependence which are very close
to those obtained for GBW), or whether it
differs from $\propto A^{1/3}$ 
due to an anomalous dimension included e.g. in the AS initial condition. 
On the other hand, running coupling evolution is seen to reduce the
$A$-dependence with increasing rapidity. We find that if fitted in a 
wide rapidity range, the dependence of $\ln{[Q_{sA}^2(Y)/
Q_{sp}^2(Y)]}$ on $Y$ is $\sim Y^{-0.4}$. However, for large values of
$A$ and $Y$, the decrease with increasing $Y$ is $\propto 1/\sqrt{Y}$ and 
thus well described by (\ref{finmue}) \cite{Mueller:2003bz}.

Combining the rescaling argument based 
in (\ref{nucres}) with the observation that the DLL solution is approached for
small $r$ or large transverse momentum $k$, one is led to an interesting
implication for the large-$k$ behaviour of the ratios of gluon
densities in nuclei over nucleon (or central over peripheral nucleus)
\cite{Albacete:2003iq,Baier:2003hr,Kharzeev:2003wz,Kharzeev:2004yx}.
In fixed coupling evolution the rescaling of the initial condition
(\ref{nucres}) trivially implies the same rescaling in the evolved solution,
which we will consider to be DLL for sufficiently large $k$.
Thus one gets for the ratio $R$ of the gluon
densities in transverse momentum space for nuclei over nucleon
\beq
R\simeq
\left[{\ln{(k^2/\Lambda^2)}-\ln{h}\over \ln{(k^2/\Lambda^2)}}\right]^{-3/4}
\,\exp{\left\{b(Y)\left[\sqrt{\ln{(k^2/\Lambda^2)}-\ln{h}}-
\sqrt{\ln{(k^2/\Lambda^2)}}\right]\right\}}
\,.
\label{cronin}
\eeq
This ratio tends very slowly to 1 for $k\to \infty$.
We have checked that the results of this formula agree with the numerical
computations in \cite{Albacete:2003iq} and thus
it provides justification to the apparent absence
of a return to the collinear limit, $R=1$ at $k \to \infty$, found
in this reference for the largest studied $k$-values.

\section{Conclusions} \label{conclusions}

The inclusion of a running coupling constant may be expected to 
account for important next-to-leading-log effects in the BK equation,
as has been previously the case for BFKL. This motivates the present
numerical study of the BK equation without impact parameter dependence.
Our main results are insensitive to details of the
implementation
of running coupling effects, the infrared regulation of the coupling 
constant and the choice of initial conditions which are evolved. 
They can be summarized as follows:

\begin{enumerate}

\item The rapidity dependence of the saturation momentum is 
much faster for fixed coupling constant than for the running
one, as observed previously 
\cite{Lublinsky:2001yi,
Braun:2003un,Rummukainen:2003ns,Golec-Biernat:2001if}.
It is well described by $Q_s^2(Y) \propto \exp{(\bar \alpha_s\, d\, Y)}$ 
for fixed coupling and by $Q_s^2(Y) \propto
\exp{(\Delta^\prime\sqrt{Y+X})}$ for running coupling.
For large rapidities, we find $d\simeq 4.57$ which is 
slightly smaller than the theoretical expectation
$d=4.88$ \cite{Iancu:2002tr,Mueller:2002zm}.
For running coupling, we find $\Delta^\prime\simeq 3.2$, slightly 
smaller than the expected value $\Delta^\prime\simeq 3.6$
\cite{Iancu:2002tr,Mueller:2002zm,Triantafyllopoulos:2002nz}. 
For a very limited region of $Y$, a fit 
to the exponential form $Q_s^2(Y) \propto \exp{(D\, Y)}$ works even
for running coupling, but it cannot account for the entire $Y$-range. 
For the fixed coupling case, we have checked the existence of
the sub-leading terms in the $Y$-dependence of the saturation scale proposed in
\cite{Mueller:2002zm,
Munier:2003vc,Munier:2003sj,Munier:2004xu}. As found in
\cite{Golec-Biernat:2004sx},
their precise determination depends
on the definition of the saturation scale.

\item For sufficiently large rapidity, the solution of the BK
equation with fixed coupling is known to show scaling
\cite{Armesto:2001fa,Lublinsky:2001bc,Albacete:2003iq}. We confirm
scaling for the running coupling case in agreement with
\cite{Braun:2003un,Rummukainen:2003ns}. The approach to the scaling solution
is faster with fixed than with running coupling.

\item As observed previously \cite{Braun:2003un} and at variance with
analytical estimates 
\cite{Iancu:2002tr,Mueller:2002zm,Triantafyllopoulos:2002nz}, the behaviour 
of $N(r)$ at small $r$ differs for the cases of fixed and running coupling.
For small $r<1/Q_s(Y)$, forms of the type 
$(r\,Q_s)^{2\gamma}\,\ln{(C\,r\,Q_s)}$
\cite{Mueller:2002zm} describe the solutions at sufficiently high 
rapidity, where $\gamma$, defined in a $Y$-independent fitting region, is
$\simeq 0.65$ for the fixed coupling constant
but $\gamma \sim 0.85$ for running coupling. These values are for the
limit $Y \to \infty$.

\item Arguments in \cite{Iancu:2002tr,Mueller:2002zm} suggest a lower
limit to the scaling window $r\,Q_s(Y)\sim \Lambda_{QCD}/Q_s(Y)$ below which
$N(r)$ returns to the perturbative double-leading-loga\-rith\-mic expression.
Remarkably, the scaling
forms (\ref{eq:f1}) proposed in
\cite{Mueller:2002zm,Triantafyllopoulos:2002nz}
give good fits to the solutions of BK even outside the scaling window,
for $r<
\Lambda/Q^2_s(Y)$, $\Lambda \sim 0.2$ GeV.
Hence, it is not possible to establish numerically the limit of the
scaling region as a deviation
from scaling.
However, the double-leading-log approximation
provides an equally good description of the numerical solution in
the $r$-region below the scaling window.

\item For fixed coupling, the scale invariance of the kernel
preserves any $A$-dependence of the initial condition during BK evolution. 
For running coupling and for very large energies and nuclear sizes, 
we have re-derived and checked numerically Equation (\ref{finmue}):
the $A$-dependence decreases with increasing rapidity like $1/\sqrt{Y}$
\cite{Mueller:2003bz}.
\end{enumerate}

The above results have been established by evolving over many orders of
magnitude in energy. Thus, any phenomenological application of these findings
has to assume that initial conditions can be fixed at (and perturbatively
evolved from) a sufficiently small energy scale for the non-linear evolution
to be effective in an experimentally accessible regime. Moreover,
phenomenology based on the BK equation will face at least some of the problems
known from applications of BFKL such as the question of whether and how to
implement kinematical cuts for gluon emission. Despite these caveats, it is
interesting to compare the numerical results found here to the general trends
in the data. A comparison of saturation-inspired parameterizations with data
on lepton-proton, lepton-nucleus and nuclear collisions at high energies
suggests a saturation scale $Q_{sA}^2 \propto A^\alpha \,\exp{(D\,Y)}$ with
$D\simeq 0.29$ \cite{Golec-Biernat:1998js} and $\alpha\simeq 4/9>1/3$
\cite{Armesto:2004ud} (for related phenomenological studies, see
\cite{Stasto:2000er,Freund:2002ux}).

Our results allow to discuss to which extent existing
data, showing geometric scaling, differ from
the asymptotic BK scaling behaviour. In particular, the strong 
$A$-dependence of the saturation scale seen in the data
indicates, at variance with the result from the BK scaling solution with
running coupling, that the properties of the initial nuclear condition have
not yet been washed out by non-linear small-$x$ evolution. The kinematic
range of the lepton-nucleus data studied in \cite{Armesto:2004ud,Freund:2002ux} is too small to test this evolution. Also, the
exponential $Y$-dependence of the saturation scale with $D \sim 0.3$ seen in
the data can appear naturally from BK evolution of reasonable initial
conditions over some units in rapidity in the running coupling case. But this
value of $D$ is not a property of the asymptotic solution for running
coupling. For fixed coupling, it can only be obtained with
unrealistically small values of the coupling constant. 

None of these facts contradicts non-linear BK evolution -- they simply 
illustrate that the evolution observed in experimental data has not yet reached
its asymptotic behaviour. To further advance our understanding of saturation
effects in QCD dynamics at high energies, both theoretical and 
experimental studies are required. In the context of the BK equation,
this requires the study of solutions under more
realistic conditions. In particular, the impact parameter dependence may have
a significant effect on the $A$-dependence of the saturation scale, a point
which we plan to study in the future. On the experimental side, 
the forward rapidity measurements at the BNL Relativistic Heavy Ion Collider
RHIC give access
to a kinematic window interesting for small-$x$ evolution studies.
These studies are at the very beginning. Also, in the near future
measurements at the CERN Large Hadron Collider
LHC will provide more stringent tests of small-$x$ 
evolution, extending the kinematic reach by at least three orders 
of magnitude further down in the momentum fraction $x$.

\vskip 0.8cm
\noindent{\bf Acknowledgments:}
We thank R.~Baier, J.~Bartels, M.~A.~Braun, E.~Iancu,
A.~H.~Mu\-eller and A.~Sabio Vera for useful
discussions and comments. We are grateful
to K.~Golec-Biernat for a
critical reading of the manuscript and many helpful remarks.
Special thanks are due to A.~Kovner, who
participated in an early stage of this work and made numerous useful
suggestions about the manuscript.
J.~L.~A. and J.~G.~M.
thank CERN Theory Division for hospitality, and acknowledge financial support
by the Ministerio de Educaci\'on y Ciencia of Spain (grant no. AP2001-3333)
and the Funda\c c\~ao para a Ci\^encia e a Tecnologia  of Portugal (contract
SFRH/BPD/12112/2003) respectively.


\begin{thebibliography}{99}

\bibitem{cargese} {\it QCD Perspectives on Hot and Dense Matter (NATO Science
Series II: Mathematics, Physics and Chemistry,
Vol. 87)},
Eds. J.-P.~Blaizot and E.~Iancu, Kluwer
2002.

\bibitem{Kuraev:1977fs}
E.~A.~Kuraev, L.~N.~Lipatov and V.~S.~Fadin,
Sov.\ Phys.\ JETP {\bf 45}, 199 (1977)
[Zh.\ Eksp.\ Teor.\ Fiz.\  {\bf 72}, 377 (1977)].

\bibitem{Balitsky:1978ic}
I.~I.~Balitsky and L.~N.~Lipatov,
Sov.\ J.\ Nucl.\ Phys.\  {\bf 28}, 822 (1978)
[Yad.\ Fiz.\  {\bf 28}, 1597 (1978)].

\bibitem{Gribov:1984tu}
L.~V.~Gribov, E.~M.~Levin and M.~G.~Ryskin,
Phys.\ Rept.\  {\bf 100}, 1 (1983).

\bibitem{Mueller:1985wy}
A.~H.~Mueller and J.~w.~Qiu,
Nucl.\ Phys.\ B {\bf 268}, 427 (1986).

\bibitem{McLerran:1993ni}
L.~D.~McLerran and R.~Venugopalan,
Phys.\ Rev.\ D {\bf 49}, 2233 (1994).

\bibitem{McLerran:1993ka}
L.~D.~McLerran and R.~Venugopalan,
Phys.\ Rev.\ D {\bf 49}, 3352 (1994).

\bibitem{McLerran:1994vd}
L.~D.~McLerran and R.~Venugopalan,
Phys.\ Rev.\ D {\bf 50}, 2225 (1994).

\bibitem{Jalilian-Marian:1996xn}
J.~Jalilian-Marian, A.~Kovner, L.~D.~McLerran and H.~Weigert,
Phys.\ Rev.\ D {\bf 55}, 5414 (1997).

\bibitem{Jalilian-Marian:1997gr}
J.~Jalilian-Marian, A.~Kovner, A.~Leonidov and H.~Weigert,
Phys.\ Rev.\ D {\bf 59}, 014014 (1999).

\bibitem{Jalilian-Marian:1998cb}
J.~Jalilian-Marian, A.~Kovner, A.~Leonidov and H.~Weigert,
Phys.\ Rev.\ D {\bf 59}, 034007 (1999)
[Erratum-ibid.\ D {\bf 59}, 099903 (1999)].

\bibitem{Kovner:1999bj}
A.~Kovner and J.~G.~Milhano,
Phys.\ Rev.\ D {\bf 61}, 014012 (2000).

\bibitem{Iancu:2000hn}
E.~Iancu, A.~Leonidov and L.~D.~McLerran,
Nucl.\ Phys.\ A {\bf 692}, 583 (2001).

\bibitem{Iancu:2001ad}
E.~Iancu, A.~Leonidov and L.~D.~McLerran,
Phys.\ Lett.\ B {\bf 510}, 133 (2001).

\bibitem{Ferreiro:2001qy}
E.~Ferreiro, E.~Iancu, A.~Leonidov and L.~McLerran,
Nucl.\ Phys.\ A {\bf 703}, 489 (2002).

\bibitem{Buchmuller:1995mr}
W.~Buchmuller and A.~Hebecker,
Nucl.\ Phys.\ B {\bf 476}, 203 (1996).

\bibitem{Ayala:1996em}
A.~L.~Ayala, M.~B.~Gay Ducati and E.~M.~Levin,
Nucl.\ Phys.\ B {\bf 493}, 305 (1997).

\bibitem{Balitsky:1995ub}
I.~Balitsky,
Nucl.\ Phys.\ B {\bf 463}, 99 (1996).

\bibitem{Weigert:2000gi}
H.~Weigert,
Nucl.\ Phys.\ A {\bf 703}, 823 (2002).

\bibitem{Kovchegov:1999yj}
Y.~V.~Kovchegov,
Phys.\ Rev.\ D {\bf 60}, 034008 (1999).

\bibitem{Nikolaev:1990ja}
N.~N.~Nikolaev and B.~G.~Zakharov,
Z.\ Phys.\ C {\bf 49}, 607 (1991).

\bibitem{Mueller:1993rr}
A.~H.~Mueller,
Nucl.\ Phys.\ B {\bf 415}, 373 (1994).

\bibitem{Mueller:1994jq}
A.~H.~Mueller and B.~Patel,
Nucl.\ Phys.\ B {\bf 425}, 471 (1994).

\bibitem{Kovner:2000pt}
A.~Kovner, J.~G.~Milhano and H.~Weigert,
Phys.\ Rev.\ D {\bf 62}, 114005 (2000).

\bibitem{Mueller:2001uk}
A.~H.~Mueller,
Phys.\ Lett.\ B {\bf 523}, 243 (2001).

\bibitem{Blaizot:2002xy}
J.~P.~Blaizot, E.~Iancu and H.~Weigert,
Nucl.\ Phys.\ A {\bf 713}, 441 (2003).

\bibitem{Iancu:2003uh}
E.~Iancu and A.~H.~Mueller,
Nucl.\ Phys.\ A {\bf 730}, 460 (2004).

\bibitem{Kovner:2001vi}
A.~Kovner and U.~A.~Wiedemann,
Phys.\ Rev.\ D {\bf 64}, 114002 (2001).

\bibitem{Braun:2000wr}
M.~Braun,
Eur.\ Phys.\ J.\ C {\bf 16}, 337 (2000).

\bibitem{Bartels:2004ef}
J.~Bartels, L.~N.~Lipatov and G.~P.~Vacca,
arXiv:hep-ph/0404110.

\bibitem{Levin:1999mw}
E.~Levin and K.~Tuchin,
Nucl.\ Phys.\ B {\bf 573}, 833 (2000).

\bibitem{Iancu:2002tr}
E.~Iancu, K.~Itakura and L.~McLerran,
Nucl.\ Phys.\ A {\bf 708}, 327 (2002).

\bibitem{Mueller:2002zm}
A.~H.~Mueller and D.~N.~Triantafyllopoulos,
Nucl.\ Phys.\ B {\bf 640}, 331 (2002).

\bibitem{Mueller:2003bz} A.~H.~Mueller,
Nucl.\ Phys.\ A {\bf 724}, 223 (2003).

\bibitem{Munier:2003vc}
S.~Munier and R.~Peschanski,
Phys.\ Rev.\ Lett.\  {\bf 91}, 232001 (2003).

\bibitem{Munier:2003sj}
S.~Munier and R.~Peschanski,
Phys.\ Rev.\ D {\bf 69}, 034008 (2004).

\bibitem{Munier:2004xu}
S.~Munier and R.~Peschanski,
Phys.\ Rev.\ D {\bf 70}, 077503 (2004).


\bibitem{Kimber:2001nm}
M.~A.~Kimber, J.~Kwiecinski and A.~D.~Martin,
Phys.\ Lett.\ B {\bf 508}, 58 (2001).

\bibitem{Armesto:2001fa}
N.~Armesto and M.~A.~Braun,
Eur.\ Phys.\ J.\ C {\bf 20}, 517 (2001).

\bibitem{Levin:2001et}
E.~Levin and M.~Lublinsky,
Nucl.\ Phys.\ A {\bf 696}, 833 (2001).

\bibitem{Lublinsky:2001bc}
M.~Lublinsky,
Eur.\ Phys.\ J.\ C {\bf 21}, 513 (2001).

\bibitem{Golec-Biernat:2001if}
K.~Golec-Biernat, L.~Motyka and A.~M.~Stasto,
Phys.\ Rev.\ D {\bf 65}, 074037 (2002).

\bibitem{Albacete:2003iq}
J.~L.~Albacete, N.~Armesto, A.~Kovner, C.~A.~Salgado and U.~A.~Wiedemann,
Phys.\ Rev.\ Lett.\  {\bf 92}, 082001 (2004).

\bibitem{Lublinsky:2001yi}
M.~Lublinsky, E.~Gotsman, E.~Levin and U.~Maor,
Nucl.\ Phys.\ A {\bf 696}, 851 (2001).

\bibitem{Triantafyllopoulos:2002nz}
D.~N.~Triantafyllopoulos,
Nucl.\ Phys.\ B {\bf 648}, 293 (2003).

\bibitem{Braun:2003un}
M.~A.~Braun,
Phys.\ Lett.\ B {\bf 576}, 115 (2003).

\bibitem{Rummukainen:2003ns}
K.~Rummukainen and H.~Weigert,
Nucl.\ Phys.\ A {\bf 739}, 183 (2004).

\bibitem{Kutak:2004ym}
K.~Kutak and A.~M.~Stasto,
arXiv:hep-ph/0408117.

\bibitem{Kovner:2001bh}
A.~Kovner and U.~A.~Wiedemann,
Phys.\ Rev.\ D {\bf 66}, 051502 (2002).

\bibitem{Golec-Biernat:2003ym}
K.~Golec-Biernat and A.~M.~Stasto,
Nucl.\ Phys.\ B {\bf 668}, 345 (2003).

\bibitem{Gotsman:2004ra}
E.~Gotsman, M.~Kozlov, E.~Levin, U.~Maor and E.~Naftali,
Nucl.\ Phys.\ A {\bf 742}, 55 (2004).

\bibitem{Iancu:2003zr}
E.~Iancu and A.~H.~Mueller,
Nucl.\ Phys.\ A {\bf 730}, 494 (2004).

\bibitem{Mueller:2004se}
A.~H.~Mueller and A.~I.~Shoshi,
Nucl.\ Phys.\ B {\bf 692}, 175 (2004).

\bibitem{Iancu:2004es}
E.~Iancu, A.~H.~Mueller and S.~Munier,
arXiv:hep-ph/0410018.

\bibitem{Stasto:2000er}
A.~M.~Stasto, K.~Golec-Biernat and J.~Kwiecinski,
Phys.\ Rev.\ Lett.\  {\bf 86}, 596 (2001).

\bibitem{Freund:2002ux}
A.~Freund, K.~Rummukainen, H.~Weigert and A.~Schafer,
Phys.\ Rev.\ Lett.\  {\bf 90}, 222002 (2003).

\bibitem{Armesto:2004ud}
N.~Armesto, C.~A.~Salgado and U.~A.~Wiedemann,
arXiv:hep-ph/0407018.

\bibitem{Gotsman:2002yy}
E.~Gotsman, E.~Levin, M.~Lublinsky and U.~Maor,
Eur.\ Phys.\ J.\ C {\bf 27}, 411 (2003).

\bibitem{Eskola:2002yc}
K.~J.~Eskola, H.~Honkanen, V.~J.~Kolhinen, J.~w.~Qiu and C.~A.~Salgado,
Nucl.\ Phys.\ B {\bf 660}, 211 (2003).

\bibitem{Armesto:2003yf}
N.~Armesto,
Acta Phys.\ Polon.\ B {\bf 35}, 213 (2004).

\bibitem{Gyulassy:2004zy}
M.~Gyulassy and L.~McLerran,
arXiv:nucl-th/0405013.

\bibitem{Kharzeev:2000ph}
D.~Kharzeev and M.~Nardi,
Phys.\ Lett.\ B {\bf 507}, 121 (2001).

\bibitem{Kharzeev:2001gp}
D.~Kharzeev and E.~Levin,
Phys.\ Lett.\ B {\bf 523}, 79 (2001).

\bibitem{Kharzeev:2004if}
D.~Kharzeev, E.~Levin and M.~Nardi,
arXiv:hep-ph/0408050.

\bibitem{Braun:2004qh}
M.~A.~Braun and C.~Pajares,
arXiv:hep-ph/0405203.

\bibitem{Braun:2004kp}
M.~A.~Braun,
Phys.\ Lett.\ B {\bf 599}, 269 (2004).

\bibitem{Kharzeev:2002pc}
D.~Kharzeev, E.~Levin and L.~McLerran,
Phys.\ Lett.\ B {\bf 561}, 93 (2003).

\bibitem{Baier:2003hr}
R.~Baier, A.~Kovner and U.~A.~Wiedemann,
Phys.\ Rev.\ D {\bf 68}, 054009 (2003).

\bibitem{Kharzeev:2003wz}
D.~Kharzeev, Y.~V.~Kovchegov and K.~Tuchin,
Phys.\ Rev.\ D {\bf 68}, 094013 (2003).

\bibitem{Kharzeev:2004yx}
D.~Kharzeev, Y.~V.~Kovchegov and K.~Tuchin,
arXiv:hep-ph/0405045.

\bibitem{Fadin:1998py}
V.~S.~Fadin and L.~N.~Lipatov,
Phys.\ Lett.\ B {\bf 429}, 127 (1998).

\bibitem{Ciafaloni:1998gs}
M.~Ciafaloni and G.~Camici,
Phys.\ Lett.\ B {\bf 430}, 349 (1998).

\bibitem{Ross:1998xw}
D.~A.~Ross,
Phys.\ Lett.\ B {\bf 431}, 161 (1998).

\bibitem{Kovchegov:1998ae}
Y.~V.~Kovchegov and A.~H.~Mueller,
Phys.\ Lett.\ B {\bf 439}, 428 (1998).

\bibitem{Armesto:1998gt}
N.~Armesto, J.~Bartels and M.~A.~Braun,
Phys.\ Lett.\ B {\bf 442}, 459 (1998).

\bibitem{Andersen:2003an}
J.~R.~Andersen and A.~Sabio Vera,
Phys.\ Lett.\ B {\bf 567}, 116 (2003).

\bibitem{Andersen:2003wy}
J.~R.~Andersen and A.~Sabio Vera,
Nucl.\ Phys.\ B {\bf 679}, 345 (2004).

\bibitem{Brodsky:1998kn}
S.~J.~Brodsky, V.~S.~Fadin, V.~T.~Kim, L.~N.~Lipatov and G.~B.~Pivovarov,
JETP Lett.\  {\bf 70}, 155 (1999).

\bibitem{Schmidt:1999mz}
C.~R.~Schmidt,
Phys.\ Rev.\ D {\bf 60}, 074003 (1999).

\bibitem{Forshaw:1999xm}
J.~R.~Forshaw, D.~A.~Ross and A.~Sabio Vera,
Phys.\ Lett.\ B {\bf 455}, 273 (1999).

\bibitem{Ciafaloni:2003kd}
M.~Ciafaloni, D.~Colferai, G.~P.~Salam and A.~M.~Stasto,
Phys.\ Lett.\ B {\bf 587}, 87 (2004).

\bibitem{Altarelli:2003hk}
G.~Altarelli, R.~D.~Ball and S.~Forte,
Nucl.\ Phys.\ B {\bf 674}, 459 (2003).

\bibitem{Khoze:2004hx}
V.~A.~Khoze, A.~D.~Martin, M.~G.~Ryskin and W.~J.~Stirling,
arXiv:hep-ph/0406135.

\bibitem{Mueller:1996hm}
A.~H.~Mueller,
Phys.\ Lett.\ B {\bf 396}, 251 (1997).

\bibitem{Babansky:2002my}
A.~Babansky and I.~Balitsky,
Phys.\ Rev.\ D {\bf 67}, 054026 (2003).

\bibitem{Lipatov:1985uk}
L.~N.~Lipatov,
Sov.\ Phys.\ JETP {\bf 63}, 904 (1986)
[Zh.\ Eksp.\ Teor.\ Fiz.\  {\bf 90}, 1536 (1986)].

\bibitem{Collins:1988ze}
J.~C.~Collins and J.~Kwiecinski,
Nucl.\ Phys.\ B {\bf 316}, 307 (1989).

\bibitem{Golec-Biernat:1998js}
K.~Golec-Biernat and M.~Wusthoff,
Phys.\ Rev.\ D {\bf 59}, 014017 (1999).

\bibitem{Chachamis:2004ab}
G.~Chachamis, M.~Lublinsky and A.~Sabio Vera,
arXiv:hep-ph/0408333.

\bibitem{Golec-Biernat:2004sx}
K.~Golec-Biernat,
arXiv:hep-ph/0408255.

\bibitem{Levin:1987xr}
E.~M.~Levin and M.~G.~Ryskin,
Yad.\ Fiz.\  {\bf 45}, 234 (1987)
[Sov.\ J.\ Nucl.\ Phys.\  {\bf 45}, 150 (1987)].

\bibitem{Levin:2001cv}
E.~Levin and K.~Tuchin,
Nucl.\ Phys.\ A {\bf 693}, 787 (2001).

\end{thebibliography}
\end{document}